\documentclass[runningheads]{llncs}
\usepackage{graphicx}
\usepackage{makecell}
\begin{document}

\title{Educational Game on  Cryptocurrency Investment: Using Microeconomic Decision Making to Understand Macroeconomics Principles\thanks{We thank Professor Sarah Jacobson for her insightful comments. We thank Tianyu Wu for his assistance in delivering the reflection survey of the educational game. We also benefited from the comments of participants at the SciEcon Seminar Spring 2021. This research is supported by the Teaching and Assessment Grant and Faculty Community for Active Learning Program, Duke Kunshan University, and National Science Foundation China for the project “Trust Mechanism Design on Blockchain: An Interdisciplinary Approach of Game Theory, Reinforcement Learning, and Human-AI Interactions,” where Luyao Zhang is the principal investigator. This research was approved by the Internal Review Board (IRB) at Duke Kunshan University. Jiasheng Zhu contributed to the research as an advisee of Prof. Luyao Zhang at Duke Kunshan University. Both Jiasheng Zhu and Luyao Zhang are also with SciEcon CIC, an NPO aiming at cultivating interdisciplinary research of both profound insights and practical impacts. We thank anonymous referees and editors at the Special Issue on Cryptocurrency of Eastern Economic Journal, one of the Palgrave Journals, part of Springer Nature, for their professional and thoughtful comments.}}
\titlerunning{Educational Game on Cryptocurrency Investment}
\authorrunning{J. Zhu and L. Zhang}

\author{Jiasheng Zhu\inst{2}\orcidID{0000-0003-2062-4550} \and
Luyao Zhang\inst{*}\inst{1}\orcidID{0000-0002-1183-2254}}
\institute{Duke Kunshan University, Jiangshu, 215316 China\\
\email{*corresponding author: lz183@duke.edu \\
Data Science Research Center \& Social Science Division\\
Duke Kunshan University}\\
\and 
Yale University, New Haven, CT 06520, USA \\
\email{Yale School of Management\\
Yale University\\
}
}
\maketitle              
\begin{abstract}
Gamification is an effective strategy for motivating and engaging users, which is grounded in business, marketing, and management by designing games in nongame contexts. Gamifying education, which consists of the design and study of educational games, is an emerging trend. However, the existing classroom games for understanding macroeconomics have weak connections to the microfoundations of individual decision-making. We design an educational game on cryptocurrency investment for understanding macroeconomic concepts in microeconomic decisions. We contribute to the literature by designing game-based learning that engages students in understanding macroeconomics in incentivized individual investment decisions. Our game can be widely implemented in online, in-person, and hybrid classrooms. We also reflect on strategies for improving the user experience for future educational game implementations. (JEL: A20, C90, D90; ACM: J.4, K.3.1, K.3.2, H.5.2)
\end{abstract}
\section{INTRODUCTION}
Gamification effectively motivates and engages users in multiple fields, including business, marketing, and management, by creating games based on nongame environments~\cite{dicheva_2014_gamification,hamari_2019_gamification}. However, gamification in education, which incorporates game design into educational contexts to enhance students’ learning motivations and outcomes, remains an emerging trend with few explorations~\cite{dichev_2017_gamifying,dicheva_2014_gamification,zeng_2020_to}. Although the literature has explored the motivational and immersive facets of gamifying educational activities, more concrete designs and implementations of such cross-border gamification have yet to be discussed~\cite{morenoger_2008_educational,silva_2019_practical,vlachopoulos_2017_the}. Moreover, previous classroom games for  understanding macroeconomics lack microeconomic foundations and are poorly connected to the  individual decision-making processes with which students more easily engage~\cite{morenoger_2008_educational}. On the other hand, individual investment games are available for classes in finance and microeconomics, but not for understanding macroeconomics~\cite{moinas_2016_the,wood_1992_the}. 
\par
This paper takes advantage of gamification and presents an educational game design on cryptocurrency investment in a micro scenario to facilitate students’ macroeconomic learning experience. Students are motivated to understand the pros and cons of macroeconomic policy and its applications in the real world through participation in the educational game. In addition, this paper also contributes to the literature on game-based learning and microeconomic decision-making. Instructors can quickly implement this game in Qualtrics\footnote{Qualtrics is a web-based software in which users can design surveys and collect answers without programming skills (Qualtrics, Provo, UT).} for online, in-person, or hybrid practices. We also designed a 7-point Likert scale questionnaire to understand students' perceived effectiveness of the game on different learning objectives.
\par
Section 2 introduces the macroeconomic concept of the rule-versus-discretion tradeoff and how emerging blockchain technology empowers us to reconstruct the concept in microeconomic decision-making. In Section 3, we present the design of the classroom. Section 4 reflects on user experience and strategies for improvements in future implementations. 

\section{Understanding Macroeconomics using Microeconomic Decisions}
In macroeconomics, the rule-versus-discretion tradeoff has long been debated~\cite{kydland_1977_rules}. Rules constrain policy-makers to pursue the same course of action in all circumstances, abiding by time consistency and generating long-term benefits. In contrast, policy-makers have wide latitude in designing the best policy to respond to events in a discretionary framework. Although the discretionary policy is more flexible, it sacrifices time consistency and might fail to realize the original policy goal. Scholars and policy-makers generally have different opinions regarding whether rules or discretion are a better policy choice~\cite{barro_1983_rules,crockett_1994_rules,fischer_1990_chapter,sayer_1981_macroeconomic}.
\par
However, these discussions mainly occurred within the public sector until blockchain enable algorithmic monetary policies~\cite{zhang2021optimal} that are rule-based, predetermined, and immutable. Moreover, many crypto exchanges allow individuals to use trading bots with hard-coded trading~\cite{fang2022cryptocurrency} based on intraday prices, transaction volumes, and other technical indicators~\cite{liu2022deciphering,liu2022cryptocurrency,zhang2022data,ao2022decentralized,zhang2022blockchain}. For example, blockchain-based applications such as Set~\cite{feng_2019_set} allow investors to choose between trading in social sets with strategies executed by humans and robot sets with hard-coded strategies. Thus, in the trending scenarios of crypto-asset investment, we are able to reconstruct the rule-versus-discretion choices in microindividual decision-making. Specifically, we design an investment game involving Ether (ETH), the native coin on the Ethereum blockchain that ranks second among all cryptocurrencies by market value~\cite{buterin_2013_ethereum}. In our game, students can either precommit to a rule-based automated trading algorithm or trade freely based on information updates subject to their own discretion. What is the tradeoff here? Similar to rule-based monetary policies, committing to an automated trading strategy achieves time consistency but sacrifices flexibility. Like discretionary monetary policies, retaining the right to trade by discretion allows flexible adjustments upon information updates but is subject to emotional investing or other irrational decisions~\cite{shiller_2016_irrational}. 

\section{Educational Game on Cryptocurrency Investment}
\begin{figure}
\includegraphics[width=\textwidth]{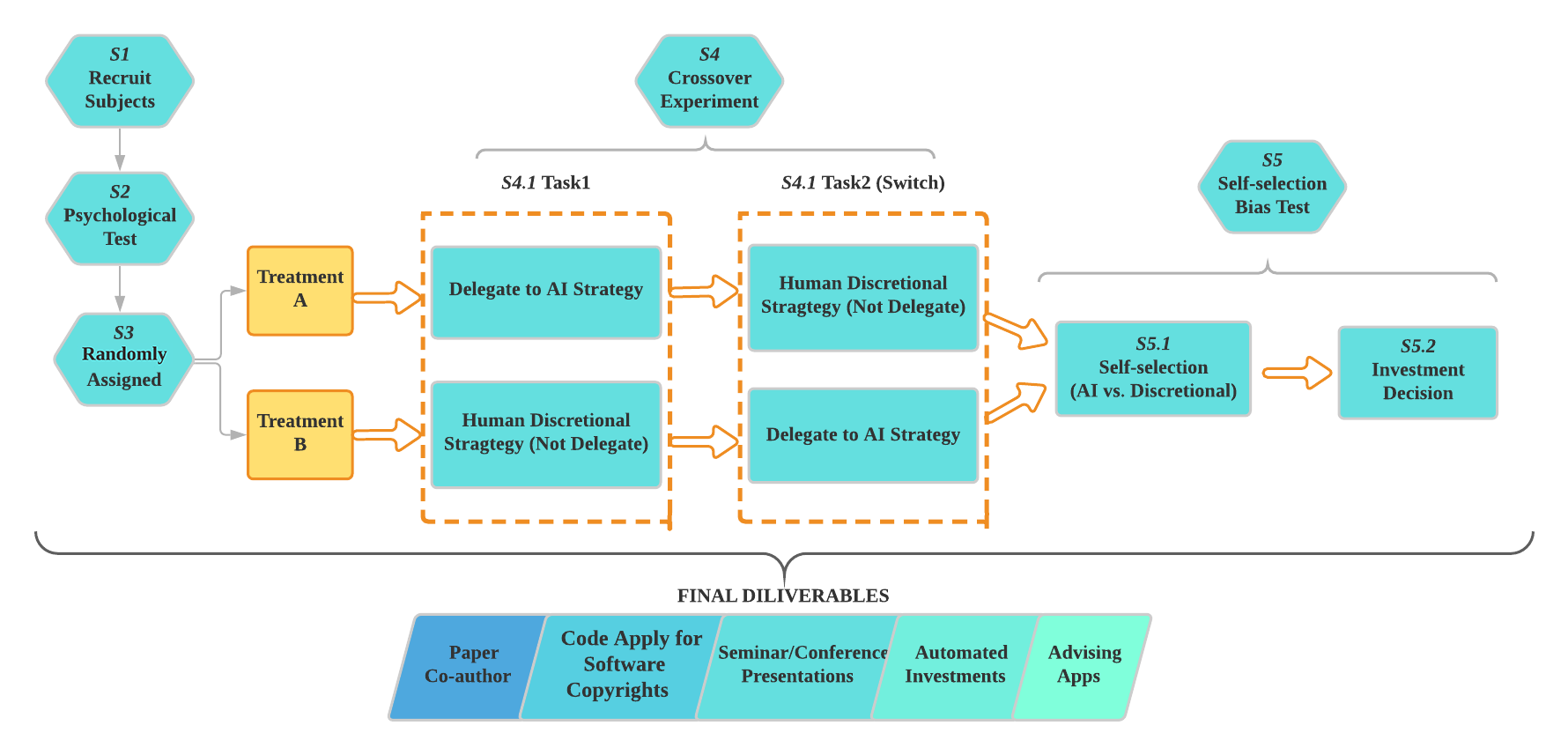}
\caption{This graph is a flow chart of the educational game design.} \label{fig2}
\end{figure}

Figure~\ref{fig2} presents an overview of the game design. We first ask the students to complete a "locus of control"~\cite{locus_of_control} psychological test. The locus of control test measures the degree to which a subject perceives outcomes resulting from internal reasons or external reasons. A higher score on the test indicates that an individual tends to attribute outcomes to external reasons rather than internal ones. Thus, our intuitive hypothesis is that people who have a higher locus of control test score are expected to favor the automated trading rule-based strategy, coined as "automated trading strategy" in the classroom game, over the human discretional strategy. 
\par
\par
Then, we randomly divide students into two treatments, A and B, with equal probabilities. Students in each treatment are asked to play in the "automated trading strategy" and "human discretional strategy" sessions sequentially but in reversed order. Finally, students are asked to play a "self-section" session where they choose to implement one of the two previous sessions that they have played.

\par

\par
We describe the game in more detail in the following 5 subsections, including the locus of control test, the introduction to the investment game, the delegation to the AI strategy session, the human discretion strategy session, and the self-selection session. Instructors may incentivize students by converting the return on investment (ROI) to extra credit, cash, or nonmonetary rewards. 
\par

\subsection{The Locus of Control Psychological Test}
\begin{quote}
Locus of control measures how strongly you perceive that you have control over the success or failure of your investment. 
Please read each statement and indicate whether it is T (true) or F (false) for yourself. There are no right or wrong answers.\\ 
\begin{itemize}
    \item 1) I usually get what I want in life.
    \item 2) I need to be kept informed about news events.
    \item 3) I never know where I stand with other people.
    \item 4) I do not truly believe in luck or chance.
    \item 5) I think that I could easily win the lottery.
    \item 6) If I do not succeed in a task, I tend to give up.
    \item 7) I usually convince others to do things my way.
    \item  8) People make a difference in controlling crime.
    \item  9) The success I have is largely a matter of chance.
    \item 10) Marriage is largely a gamble for most people.
    \item 11) People must be the master of their own fate.
    \item 12) It is not important for me to vote.
    \item 13) My life appears to be a series of random events.
    \item 14) I never try anything about which I am not sure.
    \item 15) I earn the respect and honors I receive.
    \item 16) A person can get rich by taking risks.
    \item 17) Leaders are successful when they work hard.
    \item 18) Persistence and hard work usually lead to success.
    \item 19) It is difficult to know who my real friends are.
    \item 20) Other people usually control my life.
\end{itemize}
\end{quote}
\subsection{Introduction to the Investment Game}
\begin{quote}
\footnotesize
\begin{itemize}
    \item \textbf{About Ether}: Ethereum is a decentralized, open-source blockchain with smart contract functionality. Ether (ETH) is the native cryptocurrency of the platform. It is the second-largest cryptocurrency according to market capitalization, after Bitcoin. Ethereum is the most actively used blockchain.
    \item \textbf{About the Investment Game}: This investment game consists of 3 sessions of "ether investments". In this game, you will have the chance to invest in the ether market by either AI strategy or by making your own decision. Your investment performance throughout the investment games will be converted to your extra credit. Your investment performance will be evaluated by return on investment (ROI).
\end{itemize}
\end{quote}
\subsection{Automated Trading Strategy Session}
\begin{figure}
\includegraphics[width=0.5\textwidth]{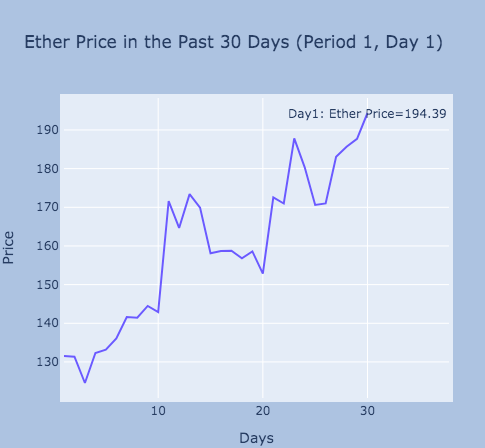}
\includegraphics[width=0.5\textwidth]{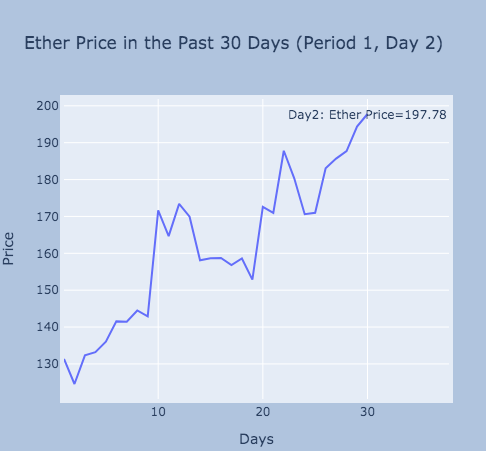}
\caption{ETH’s price in the past 30 days shown to students for investment decisions: on the first day (left) and on the second day (right).} \label{fig1}
\end{figure}

In this session, students are asked to play three periods of a 10-day (round) investment game. In each period, the ETH prices are sampled from actual historical records by a random draw of the starting date. We first present students with the ETH price in the past 30 days, as in Figure~\ref{fig1} (left). Then, we ask students to choose one of the three automated trading strategies. 
\begin{quote}
\begin{itemize}
    \item Hello, crypto investors! You are randomly assigned to the automated trading strategy mode for your round 1 ether investment. No additional operations are needed. All you need to do is click on the button and move forward until the end of the investment.
    \item You can choose one of the following three automated trading strategies:\\
    \textbf{A. Long Strategy:} buy 10 ETH in each trading day\\
    \textbf{B. Holding Strategy:} maintain the status quo and keep holding your current assets\\
    \textbf{C. Short Strategy:} sell 10 ETH in each trading day
\end{itemize}
\end{quote}
Finally, the investment results are implemented immediately by simulating the AI strategy that the student chooses for 10 rounds.
\begin{quote}
\begin{itemize}
    \item Your Initial Endowment\\
    \textbf{USD Account:} \$ 20507.6\\
    \textbf{ETH Account:} 100 ETH
    \item This graph shows the ETH price in the past thirty days.
    \item Please choose one from the following three automated trading strategies for the first investment period (the following ten days):\\
    \textbf{A. Long Strategy:} buy 10 ETH in each trading day\\
    \textbf{B. Holding Strategy:} maintain the status quo and keep holding your current assets\\
    \textbf{C. Short Strategy:} sell 10 ETH in each trading day
\end{itemize}
\end{quote}
\subsection{Human Discretion Strategy Session}
As in the "automated trading strategy" session, students are asked to play three periods of a 10-day (round) investment game. In each period, the ETH prices are sampled from actual historical records by a random draw of the starting date. Unlike the "automated trading strategy" session, students in total make decisions for ten days (rounds) instead of one precommitment decision. Moreover, we present to students the updated price chart for the past 30 days before they make their decisions for each day (round), shown as in Figure~\ref{fig1} (left and right).
\begin{quote}
\begin{itemize}
    \item Your Initial Endowment\\
    \textbf{USD Account:} \$ 20507.6\\
    \textbf{ETH Account:} 100 ETH
    \item Today is Day 1 in your first investment period. This graph shows ETH prices in the past 30 days.
    \item Please make your investment decision by clicking one of the following options:\\
    A. Buy 10 ETH\\
    B. Hold\\
    C. Sell 10 ETH
\end{itemize}
\end{quote}

\subsection{The Self-Selection Session}
Students are asked to play another three periods of a 10-day (round) investment game. Different from previous sessions, students can now choose between an "automated trading strategy" session and a "human discretional strategy" session. 

\par
At the end of the game, the instructor can guide students' reflections. For example, the instructor may ask the following open-ended questions:
\begin{quote}
\begin{itemize}
    \item \footnotesize A. Elicit one research question that the experimental design and results can answer (test). Is there any difference between your intuitive choice and what economic theory dictates? 
    For instance, below are the three types of research questions summarized from the post-experiment reflections and research proposals of the students who attended our classroom game. 
    \begin{itemize}
    \item \footnotesize a. Performance Study: Conduct cross-subject tests on the ROI of automated trading and discretion strategies. Which, "automated trading strategy" or "human discretion strategy", has a better investment performance?
    \item \footnotesize b. Rationality Study: Are subjects' decisions in the "self-selection session" consistent with the better-performing strategy used in the previous sessions? Are subjects rational?
    \item \footnotesize c. Behavioral Study: Compute the correlations between the locus of control test scores and a dummy variable (1 for the "automated trading strategy" and 0 for the "human discretion strategy"). Can the subjects' decision-making be explained by the locus of control test?
    \end{itemize}
    
    \item \footnotesize B. How can policymakers or mechanism designers help individuals make better investment decisions based on the rule-versus-discretion trade-off? For example, below are three facets summarized from the post-experiment reflections in which the students think policymakers or mechanism designers could further explore to help individuals make evaluate the rule-versus-discretion trade-off. 
    
    \begin{itemize}
    \item \footnotesize a. Attention Cost: Since the "human discretion strategy" requires more attention from subjects to make the daily investment decisions one by one, is there a discretion cost compared to the "automated trading strategy"?
    \item \footnotesize b. Flexibility Cost: Subjects who play the "human discretion strategy" maintain flexibility in regard to choosing sell, buy, or hold decisions while those who play with the "automated trading strategy" forgo this right. Is there a flexibility cost in this trade-off?
    \item \footnotesize c. Emotion Investment: Investment decisions are made either due to the change in market fundamentals or investors' emotions. If the market fundamentals remain the same, does the "human discretion strategy" lead to emotional investment (e.g. buying high and selling low) that the "automated trading strategy" can avoid?
    \end{itemize}

\end{itemize}
\end{quote}
These reflection questions guide students to think from the perspectives of both a policy-maker and a decision-maker. Thus, the reflecting exercises further inspire students to research new macro policy solutions to the rule-versus-discretion dilemma in a microeconomic decision-making scenario. The instructor can also compare the 3 test results from Research Question A to help students better meet learning goals. Each of the students is required to submit a research proposal on rule-versus-discretion as their final project in the course.
\par
Students' understanding of the rule-versus-discretion trade-off can be assessed according to the sample questions we list under Research Question A. In addition, we design a survey to evaluate their knowledge absorption situation after the game. Instructors can also simply chat with students to obtain oral feedback about what they have learned from the game.

\subsection{Frequently Asked Questions (FAQ)}
This subsection lists several frequently asked questions about the game. We can use this list to answer students' questions in the actual implementation. 
\par
\begin{quote}
\begin{itemize}
    \item \footnotesize A. Question: Are the ether investments sessions independent? Do the students keep their ethers from the previous sessions? Are the students’ earnings added to their USD account after each session? 
    \par
    -- Answer: Yes, the sessions are independent. Students do not keep ethers from the previous session; and thus, their earnings are not compounded.
    
    \item \footnotesize B. Question: Do the students in the rule-based sessions see the change in the ETH account and USD account after each round or only once at the end of the session? What about the students in the discretionary treatment? 
    \par
    -- Answer: In either session, students only see the change in ETH account and USD account at the end of the session.
    
    \item \footnotesize C. Question: How frequently does the random drawing of the starting date occur: between each session or only once for each student?
    \par
    -- Answer: The random drawing of the starting date occurs when a new session begins.
    
    \item \footnotesize D. Question: Does committing to automated trading mean that students forgo any control? Are the students made aware of what automated trading does in its algorithm?
    \par
    -- Answer: Yes, committing to automated trading means that students forgo any control of their investment decision in one period of a 10-day (round) investment game. One can only choose another automated trading strategy until the current period of a 10-day (round) investment game ends. Students are given an introduction to automated trading strategies before the game; thus, they are aware of what automated trading does in its algorithm in our design.
    
\end{itemize}
\end{quote}

\section{Reflection on Gamification Implementation}
We implemented a pilot game in the Intermediate Macroeconomics course at Duke Kunshan University in the Spring of 2021. The prerequisites for this course include an introductory-level economics course and calculus. Before the game started, we introduced students to the rule-versus-discretion dilemma in macroeconomics. To test the user experience, we asked students to complete a 7-point Likert scale questionnaire regarding their perceived engagement in the game and the perceived game effectiveness on 7 different learning objectives: engagement, remembering concepts, understanding different perspectives on macroeconomic issues, analyzing real-world issues, collaborative learning, conducting systematic research, and understanding the pros and cons of macroeconomic policy.~\footnote{We refer the readers to Jacobson et. al.~\cite{jacobson2022right} for a more comprehensive survey evaluating the effectiveness of different active learning techniques on different learning objectives.} Table 1 shows that 75\% of the students at least somewhat agreed that they felt engaged in the game. In general, the majority thought that the game effectively achieves the six learning objectives. However, the distributions of answers are different across learning objectives. For example, 67.86\% of students, at least, somewhat agreed that the game effectively analyzes real-world issues with macro theories; however, only half of the students, at least, somewhat agreed on the effectiveness for remembering macroeconomics concepts.

\begin{table}[htbp]
\footnotesize
\centering
\caption{Summary of the user experience questionnaire responses: the students' perceived engagement and effectiveness on various learning objectives }\label{tab1}
\begin{tabular}{|l|c|c|}
\hline
\thead{\makecell{Questions for \\Learning Objectives }}
 &  \thead{\makecell{\footnotesize somewhat agree\\very agree\\extremely agree \\(percent)}} &  \thead{\makecell{\footnotesize neither disagree nor agree\\somewhat disagree\\very disagree\\extremely disagree \\(percent)}}\\
\hline
{\thead{\makecell{Q1. engagement in the educational game}}} &  {75.00\%} & {25.00\%} \\
\thead{\makecell{Q2. effectiveness in remembering \\concepts in macroeconomics}} &  50.00\% & 50.00\% \\
\thead{\makecell{Q3. effectiveness in understanding \\perspectives on macroeconomic issues}} &  60.71\% & 39.29\% \\
\thead{\makecell{Q4. effectiveness in analyzing real-world\\ issues with macro theories}} &  67.86\% & 32.14\% \\
\thead{\makecell{Q5. effectiveness in collaborative learning}} & 57.14\% & 42.86\% \\
\thead{\makecell{Q6. effectiveness in conducting \\systematic research}} &  64.29\% & 35.71\% \\
\thead{\makecell{Q7. effectiveness in understanding \\the pros \& cons of macroeconomic policy}} &  64.29\% & 35.71\% \\
\hline
\end{tabular}
\end{table}
\par
The game process lasted less than one hour, which means that it is convenient for instructors to implement during class time. Moreover, the game was implemented in a hybrid classroom, which means that it is applicable to online and in-person teaching as well. As of this writing, the most recent pandemic is still impacting education. Thus, we envision the vast potential that the educational game has in connecting faculty and students to achieve various learning objectives in remote education. Before the game, most students had not heard of the rule-versus-discretion trade-off. However, after the game, students expressed in post-experiment discussions that they had a much deeper understanding of the topic and had realized how important it could be. All of them completed a research proposal on rule-versus-discretion independently as a part of their course assignments.
\par
However, we can identify one major limitation of the current implementation, namely, the user interface is much less interactive than commercialized investment applications. To improve the user experience, we see the necessity of deploying blockchain applications for educational games. Although the game industry has been blooming on blockchains recently~\cite{cryptogames2019,cryptokitties2021}, the blockchain applications for educational games still need to be designed and deployed~\cite{zhang2023design}.  

\bibliographystyle{splncs04}
\footnotesize
\bibliography{citation}

\end{document}